# Specular Andreev reflection of asymmetric fermions in graphene under strain

**Bumned Soodchomshom**

Thailand Center of Excellence in Physics, Commission Higher on Education, Ministry of Education, Bangkok 10400, Thailad

**Abstract**

This work investigates the effect of the uniaxial strain on the tunneling conductance in a strained graphene superconductor where strain is applied in the armchair direction. Based on the Tight-Binding model, applying strain in the armchair direction gives rise to the asymmetric massless fermions as the carriers. Their velocities depend on their directions controlled by strain. Using the BTK theory, the conductances of strained graphene N/S junctions can be determined. As a result, we find that the current flowing perpendicular to the direction of strain depends linearly on strain, with the positive slope. But the current flowing parallel to the direction of strain depends linearly on strain, with the negative slope. This linear behavior is significant for applications of superconductor-based nanomechanical electronic devices.

## 1. Introduction

Since graphene, a monolayer of graphite, has been discovered [1], it became a great topic of interest in the field of the condensed matter. One of the famous properties of graphene is its carriers behaving like the massless relativistic particles [2, 3]. This has lead to several interesting phenomena which are not observed in the conventional condensed system. The bridge between relativistic and superconducting natures was first done in the graphene system. The novel Cooper pairs are formed by the relativistic massless electrons with momentum k and spin up and the massless Dirac electrons with momentum –k and spin down. Hole reflection or the Andreev reflection due to the tunneling of the quasielectron from normal graphene (N) into superconducting graphene (S) was first theoretically studied by Beenaker [4]. The specular Andreev reflection occurs when the Fermi energy in N is smaller than the excited energy. This behavior results from the specific nature of the Cooper pairs formed by the massless relativistic particles. The tunneling property in graphene-based superconductor system has been, next, extended to study in several junctions, for instance, N/S, F/S, N/gate-barrier (NB)/S and N/magnetic barrier (FB)/S structures [5-10]. Also, supercurrents in S/NB/S, S/FB/S and S/FB/FB/S junctions have been investigated [11-15].

Recently much interest has been given to the interplay between mechanical and electronic properties in graphene [16-22]. As the important knowledge for application of graphene-based nanomechanical electronics, the effect of strain on the electronic property of graphene under tension was investigated in both experiments [16-19] and theories [20-23]. A uniaxial strained graphene can be performed by depositing graphene on the top of the flexible sheet of polyethylene terephthalate and then stretching polyethylene terephthalate sheet in one direction [16]. Also, applying tension in a graphene sheet gives rise to the deformed honeycomb lattice. The asymmetric graphene structure because of strain leads electrons in the sublattices asymmetrically interacting to the three nearest neighbor electrons. The three hoping energies and the three positions of the nearest neighbor electrons are altered differently. Due to this, the effective energy band structure becomes asymmetric. The velocities of the carriers, consequently, depend on their directions. Graphene is gapless as governed by the asymmetric massless fermions since strain is applied in the armchair direction [22, 23], unlike that in the case of strain applying in the zigzag direction. Gapped graphene may be created at the critical deformation [22, 23]. As the best property, controllable electronic property of graphene by strain leads to the connection of the mechanical property and the electronic property. The limit of strain to deform the honeycomb lattice in graphene by applying tension is that large tension may permanently breaks its structure. Strain which is larger than 20 % may break the honeycomb structure of graphene [20]. Because of this limit, tuning hoping energies in the graphene system by tension must be modeled under strain lower than 20%. Recently, simulation of the effect of uniaxial strain on the transport property of a nano-ribbon graphene-based tunneling field effect transistor was investigated [21]. This is importance for an application of the nano-mechanical electronic devices, field effect transistor.

In this paper, we investigate the tunneling conductance in a strained graphene N/S junction where the strain is applied along the **armchair direction**, y-direction (see Fig.1a). Applying strain in this direction gives rise the **asymmetric massless fermions** as the carriers. Graphene is gapless for all values of applied strain [22, 23]. Unlike applying strain in the zigzag direction, gapless graphene may turn to gapped graphene at the critical deformation (see the discussion in refs.22 and 23). In this work, the Hamiltonian for the carriers in the strained graphene system is obtained, based on the Tight-Binding model. Using the Dirac Bogoliubov-de Gennes equation (DBdG), we next consider the two current types; (i) the current $I_x$ flowing perpendicular to the direction of strain (or tension), seen in Fig.1b and (ii) the current $I_y$ flowing parallel to the direction of strain, seen in Fig.1c. The conductances $G_{x(y)}$ related to the currents $I_{x(y)}$ are calculated by using the BTK formalism [24]. In particular, the present work shows the important fundamental mechanical-electronic property in a graphene superconductor-based system.

## 2. Model of the free particle Hamiltonian for graphene under strain

We first consider the mechanical effect on electronic property based on the Tight-Binding model. The Hamiltonian is, therefore, given by

$$\hat{H} = -t_{ij} \sum_{<i,j>} (a_i^* b_j + H.c) , \qquad (1)$$

where <i,j> represents the nearest neighbor sites, and $a_i$ ( $b_j$ ) is annihilation operator at sublattice A ( B) and $t_{ij}$ are the hoping energies related to the three vector displacements $\vec{\sigma}_{1,2,3}$ as seen in Fig.1a. When graphene is under tension, the vector displacements are changed as a function of strain S. These are obtained, in this model, by the formulae $\vec{\sigma}_1 = c_0([\sqrt{3}/2](1-pS)\hat{x} + [1/2](1+S)\hat{y})$, $\vec{\sigma}_2 = c_0([-\sqrt{3}/2](1-pS)\hat{x} + [1/2](1+S)\hat{y})$ and $\vec{\sigma}_3 = -c_0(1+S)\hat{y}$, where $\hat{x}$ and $\hat{y}$ are the unit vectors along the +x and the +y directions, respectively. The deformed angle $\Theta$ can be obtained by $\Theta = \mathrm{Tan}^{-1}[\sqrt{3}(1-pS)/(1+S)]$. In this model, $c_0$ =0.142 $A^0$ and $p \cong$ 0.165 [21] are c-c distance in unstrained graphene and Poisson’s ratio assumed as for graphite, respectively.

The effective Hamiltonian for electron field with wave vector $\vec{k} = k_x\hat{x} + k_y\hat{y}$ in the uniaxial strained graphene related to eq.(1), is given by

$$\hat{H}_{eff} = \begin{bmatrix} 0 & -t_s \sum_{s=1}^{3} \exp[i\vec{k}.\vec{\sigma}_s] \\ -t_s \sum_{s=1}^{3} \exp[-i\vec{k}.\vec{\sigma}_s] & 0 \end{bmatrix}, \quad (2)$$

where $t_{1,2,3}$ which are, in this model, defined as of the form $t_s = t_0 e^{-\beta(|\vec{\sigma}_s|/c_0-1)}$ with $t_0$ being the hoping energy in the equilibrium graphene and $\beta \cong 3.37$ is assumed [22]. In this model, $t_1$=$t_2$=t and $t_3$=$t'$. The Eigen energy suited for the Hamiltonian in eq.(2) is obtained as

$$E_k = \pm t' \sqrt{1 + 4(\frac{t}{t'})^2 \cos^2[k_x L \sin\Theta] + 4(\frac{t}{t'}) \cos[k_x L \sin\Theta] \cos[k_y \{L\cos\Theta + L'\}]} \, . \quad (3)$$

Here, in the case of unstrained graphene, $t = t' = t_0 \cong 2.7\text{eV}$ [25], $S = 0$ and $\Theta = 60^0$.

By using eq.(2) and (3), the Hamiltonian and the energy spectrum expanded around the Dirac point $(k_D, 0) \sim (\frac{1}{L_x}\cos^{-1}\left[\frac{-t'}{2t}\right], 0)$ can be obtained as

$$\hat{H}_{eff} \cong \hbar \begin{bmatrix} 0 & v_x q_x - i v_y q_y \\ v_x q_x + i v_y q_y & 0 \end{bmatrix},$$

and $$E_k \cong \pm\hbar\sqrt{v_x^2 q_x^2 + v_y^2 q_y^2}\,, \quad (4)$$

respectively, where $q_x = k_x - k_D$ and $q_y = k_y$.

$$v_x = 2tL\sin[\Theta]\sin[\cos^{-1}[\frac{-t'}{2t}]]/\hbar\,,$$

and $$v_y = t'(L\cos\Theta + L')/\hbar\,. \quad (5)$$

The parameters, $L\sin\Theta = \frac{\sqrt{3}}{2}c_0(1-pS)$, $L\cos\Theta = \frac{1}{2}c_0(1+S)$, $L' = c(1+S)$ and $v_F = \frac{3t_0c_0}{2\hbar} \sim 10^6$ m/s are applied to this model. We have given the result that the carriers in strained graphene are asymmetric massless Dirac fermions. Their velocities depend on their directions, unlike those in the unstrained graphene [2, 3]. The Model of our Hamiltonian in eq.4 agrees with the Hamiltonian given by ref.23 for the case of strain applied in the armchair direction.

## 3. Scattering process and conductance formalism

To consider the scattering problem at the interface N/S, we first study in the case of the current flowing along the x-direction. Since the Cooper pairs in the strained graphene are formed by the asymmetric massless Dirac electrons (see eqs. 4 and 5) with momentum k and spin up and the asymmetric massless Dirac electrons with momentum –k and spin down, the motions for the quasiparticles with the exited energy E are governed by the asymmetric DBdG equation. We then have

$$\begin{pmatrix} -i\hbar(\sigma_x v_x \partial_x + \sigma_y v_y \partial_y) - E_F(x) & \Delta(x) \\ \Delta^*(x) & i\hbar(\sigma_x v_x \partial_x + \sigma_y v_y \partial_y) + E_F(x) \end{pmatrix} \psi(x)e^{ik_y y}$$
$$= E\psi(x)e^{ik_y y} \qquad (6)$$

where $\sigma_{x,y}$ are the Pauli spin matrices related to the x- and y- directions. We define the Fermi energy in the system as $E_F(x<0)=E_F$ and $E_F(x>0)=E_F+u$ and also define the superconducting order parameter as $\Delta(x<0)=0$ and $\Delta(x>0)=\Delta e^{i\phi}$ with $\phi$ being the superconducting phase. For the current flowing along the x-direction, the conservation momentum is $k_y=k_{//}$. The solution for the wave function in each region is

, in the N region,

$$\psi(x<0)=\begin{pmatrix}1\\A_{e+}\\0\\0\end{pmatrix}e^{ik_{eN}\cos\theta x}+b\begin{pmatrix}1\\A_{e-}\\0\\0\end{pmatrix}e^{-ik_{eN}\cos\theta x}+a\begin{pmatrix}0\\0\\1\\A_h\end{pmatrix}e^{ik_{hN}\cos\theta_A x},$$

and, in the S region,

$$\psi(x>0)=c\begin{pmatrix}1\\C_e\\e^{-i(\upsilon+\phi)}\\C_e e^{-i(\upsilon+\phi)}\end{pmatrix}e^{ik_{eS}\cos\theta_{eS}x}+d\begin{pmatrix}1\\C_h\\e^{-i(-\upsilon+\phi)}\\C_h e^{-i(-\upsilon+\phi)}\end{pmatrix}e^{-ik_{hS}\cos\theta_{hS}x},$$

(7)

where $A_{e\pm}=\dfrac{E_F+E}{\hbar(\pm v_x k_{eN}\cos\theta-iv_y k_{//})}$ , $A_h=\dfrac{E_F-E}{\hbar(v_x k_{hN}\cos\theta-iv_y k_{//})}$,

$$C_e=\frac{E_F+u+\sqrt{E^2-\Delta^2}}{\hbar(v_x k_{eS}\cos\theta_{eS}-iv_y k_{//})}\ ,\ C_h=\frac{E_F+u-\sqrt{E^2-\Delta^2}}{\hbar(-v_x k_{hS}\cos\theta_{hS}-iv_y k_{//})},$$

with $$k_{e(h)N}=\frac{E_F+(-)E}{\hbar(\sqrt{(v_x\cos\theta_{(A)N})^2+(v_y\sin\theta_{(A)N})^2}}\ ,$$

$$k_{e(h)S}=\frac{E_F+u+(-)\sqrt{E^2-\Delta^2}}{\hbar(\sqrt{(v_x\cos\theta_{e(h)S})^2+(v_y\sin\theta_{e(h)S})^2})},$$

and $e^{i\upsilon}=\dfrac{E}{|\Delta|}+\sqrt{\left(\dfrac{E}{\Delta}\right)^2-1}$ . (8)

$\theta$ and $\theta_A$ are the injected angles of electron and the incident angle of hole in N, respectively. Also, $\theta_{eS}$ and $\theta_{hS}$ are the incident angles of electron and hole in S. These angles can be calculated through the conservation condition. The coefficients a, b, c and d can be determined by matching the wave function in eq.(7) with the boundary condition at x=0, $\psi(0^-)=\psi(0^+)$. We then have the Andreev reflection and the normal reflection coefficients, as given by

$$a = \frac{e^{i(\upsilon-\phi)}(A_{e+} - A_{e-})(C_e - C_h)}{(A_h - C_e)(A_{e-} - C_h) + e^{2i\upsilon}(C_e - A_{e-})(A_h - C_h)},$$

and
$$b = \frac{(C_e - A_h)(A_{e+} - C_h) + e^{2i\upsilon}(C_e - A_{e+})(C_h - A_h)}{(A_h - C_e)(A_{e-} - C_h) + e^{2i\upsilon}(C_e - A_{e-})(A_h - C_h)},$$

(9)

respectively.

This work calculates the conductance $G_x$ related to the current $I_x$ by using the BTK formalism [24]. The dimensionless conductance is, therefore, given by

$$G_x \sim \int_0^{\theta_C} d\theta \cos\theta (1 + \frac{\cos\theta_A}{\cos\theta}|a|^2 - |b|^2) \ ,$$

where $\theta_C = \cot^{-1}\left(\frac{v_x}{v_y}\sqrt{\left(\frac{E_F + E}{E_F - E}\right)^2 - 1}\right)$ . (10)

In the case of $G_y$, the conductance related to the current $I_y$, it is determined easily by inter change $v_x \leftrightarrow v_y$ in the previous formulae.

## 4. Result and discussion

Let us first consider the numerical result showing the effect of strain S, in the case of current flowing along the x-direction, on the plotted conductance $G_x$ versus the biased voltage V. As seen in Figs.2a-c, we set the Fermi energy in N, $E_F \sim 0.2\Delta$ in order to take into account the effect of the retro and the specular Andreev reflections on the conductance. In the case of weakly doped graphene in S (see Fig1a), we also approximately set $u \sim 5\Delta$. We find that, the conductance increases when increasing strain for all biased voltage V. To increase u up to $1000\Delta$ for a heavily doped graphene in S (see Fig.1b), it is shown that the result is found to be the same as that in the weakly doped graphene in S. This can be concluded that the effect of strain on the conductance does not depend on u. Interestingly, for the large Fermi energy $E_F$ shown in Fig.2c the conductance does not depend on strain. This prediction may be easily tested experimentally.

We next study the effect of strain on the conductance $G_y$ (see Figs.3a-c). In this case, the current we consider flows parallel to the direction of strain. A given result shows very different from $G_x$. The conductance exhibits decreasing for increasing strain. This difference is due to the asymmetric nature of the carriers that their velocities depend on their directions. However, as is similar to $G_x$, we also find that the conductance does not depend on u and does not depend on strain for large $E_F$.

In Figs.4a-b, we show the interesting result may be useful for the applications of strain-control-current devices, a superconductor-based nanomechanical electronics. The conductances are plotted as a function of strain, at the zero bias and for the weakly doped graphene superconductor. Perfectly, the conductances $G_x$ and $G_y$ depend linearly on strain. The slopes of the linear curves are positive and negative for $G_x$ and $G_y$, respectively. And they can also be tuned by varying $E_F$. The conductances strongly depend on strain for small $E_F$.

## 5. Summary

This work has investigated the effect of the uniaxial strain on the conductance of a N/S junction where strain is applied in the armchair direction. Based on the Tight-Binding model, we have obtained the asymmetric massless Dirac fermions as the carriers of the strained graphene system. As an important result, we found that the current flowing perpendicular to the direction of strain depends linearly on strain, with the positive slope. But the current flowing parallel to the direction of strain depends linearly on strain, with the negative slope. This linear behavior is valuable for superconductor-based nanomechanical electronic applications of strain-control-current devices.

## Figure captions

**Figure 1.** schematic illustrations of (a) the microscopic structure of the deformed honeycomb lattice when applying tension along the y-direction, (b) the current $I_x$ flowing perpendicular to the direction of strain (or tension) and (ii) the current $I_y$ flowing parallel to the direction of strain.

**Figure 2.** conductance $G_x$ related to $I_x$ is studied as a function of biased voltage V for strain s=0, 0.1 and 0.2, (a) for $E_F = 0.2\Delta$ and $u = 5\Delta$, (b) for $E_F = 0.2\Delta$ and $u = 1000\Delta$, and (c) for $E_F = 1000\Delta$ and $u = 0$.

**Figure 3.** conductance $G_y$ related to $I_y$ is studied as a function of biased voltage V for strain s=0, 0.1 and 0.2, (a) for $E_F = 0.2\Delta$ and $u = 5\Delta$, (b) for $E_F = 0.2\Delta$ and $u = 1000\Delta$, and (c) for $E_F = 1000\Delta$ and $u = 0$.

**Figure 4.** the conductances plotted as a function of strain at zero bias and $u = 5\Delta$ for various Fermi energies, (a) $G_x$, and (b) $G_y$.

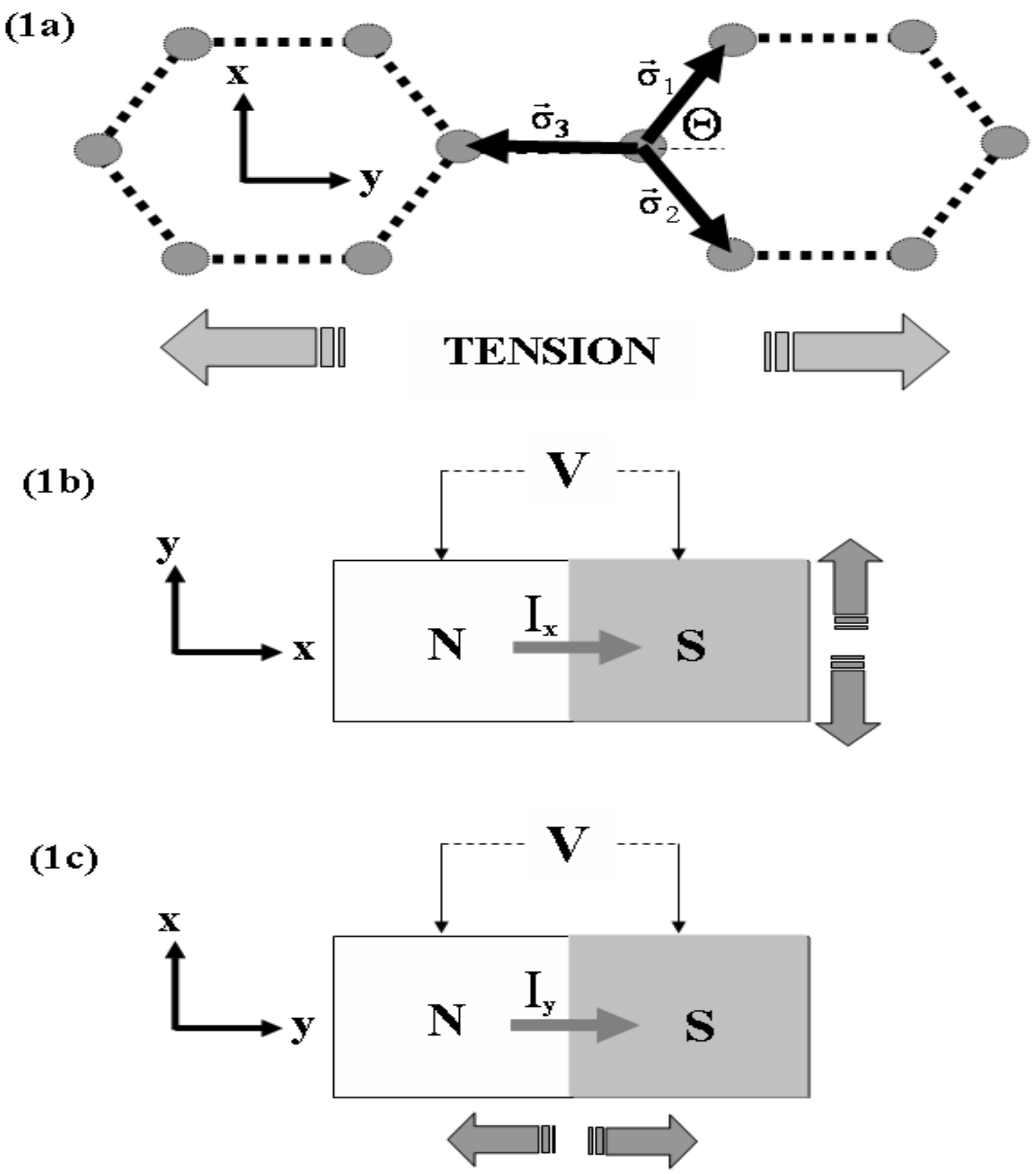

**Figure 1**

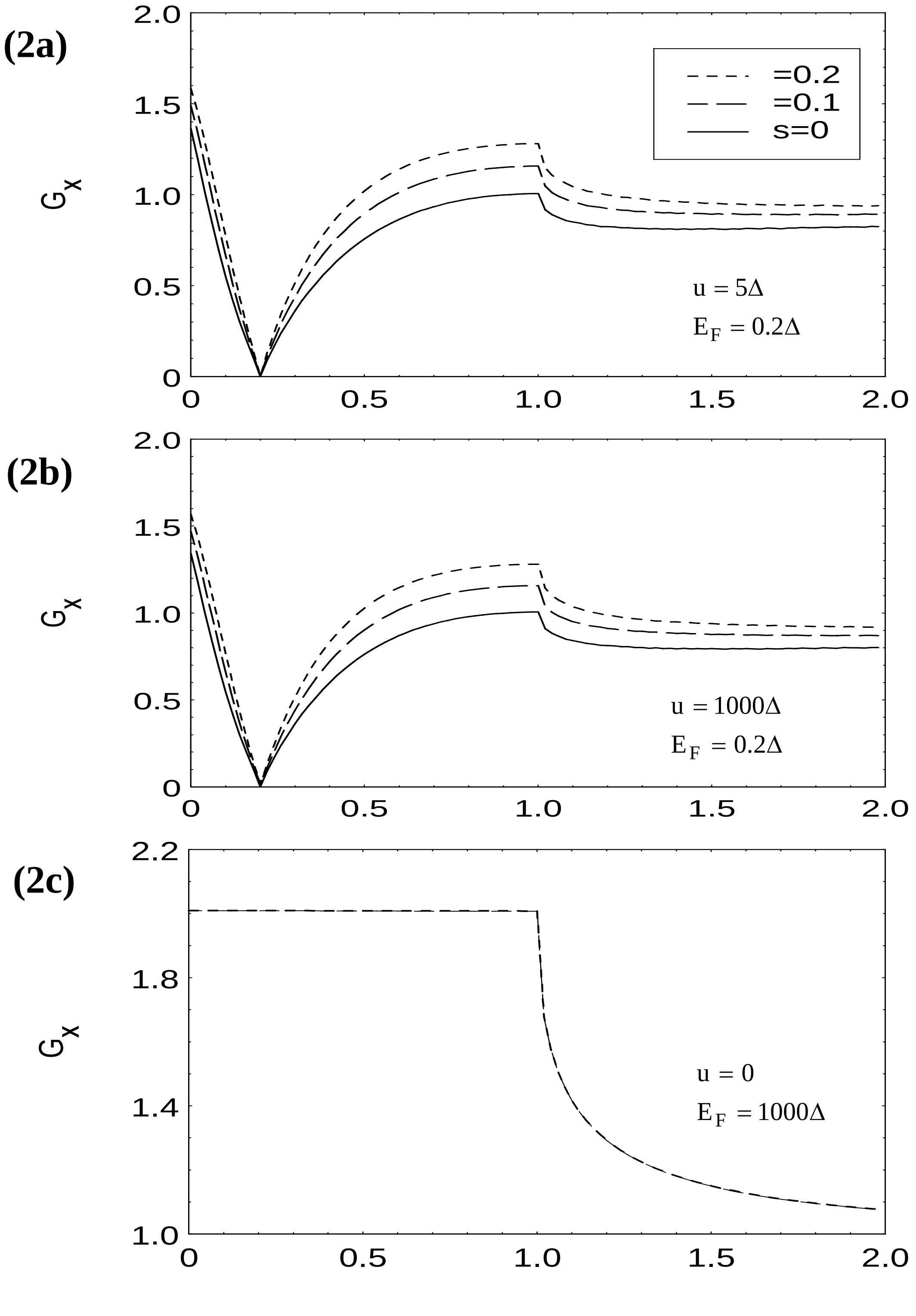

**Figure 2**

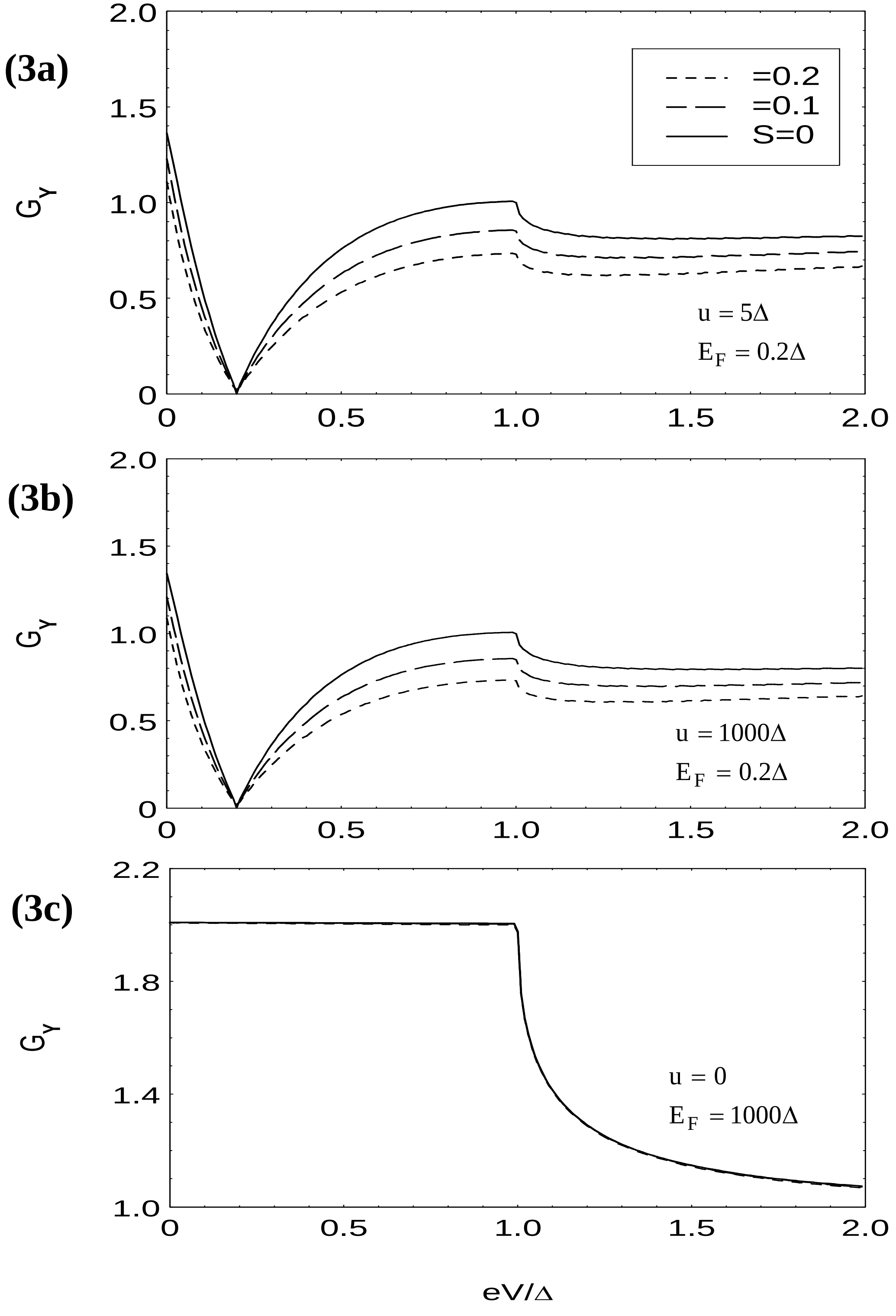

**Figure 3**

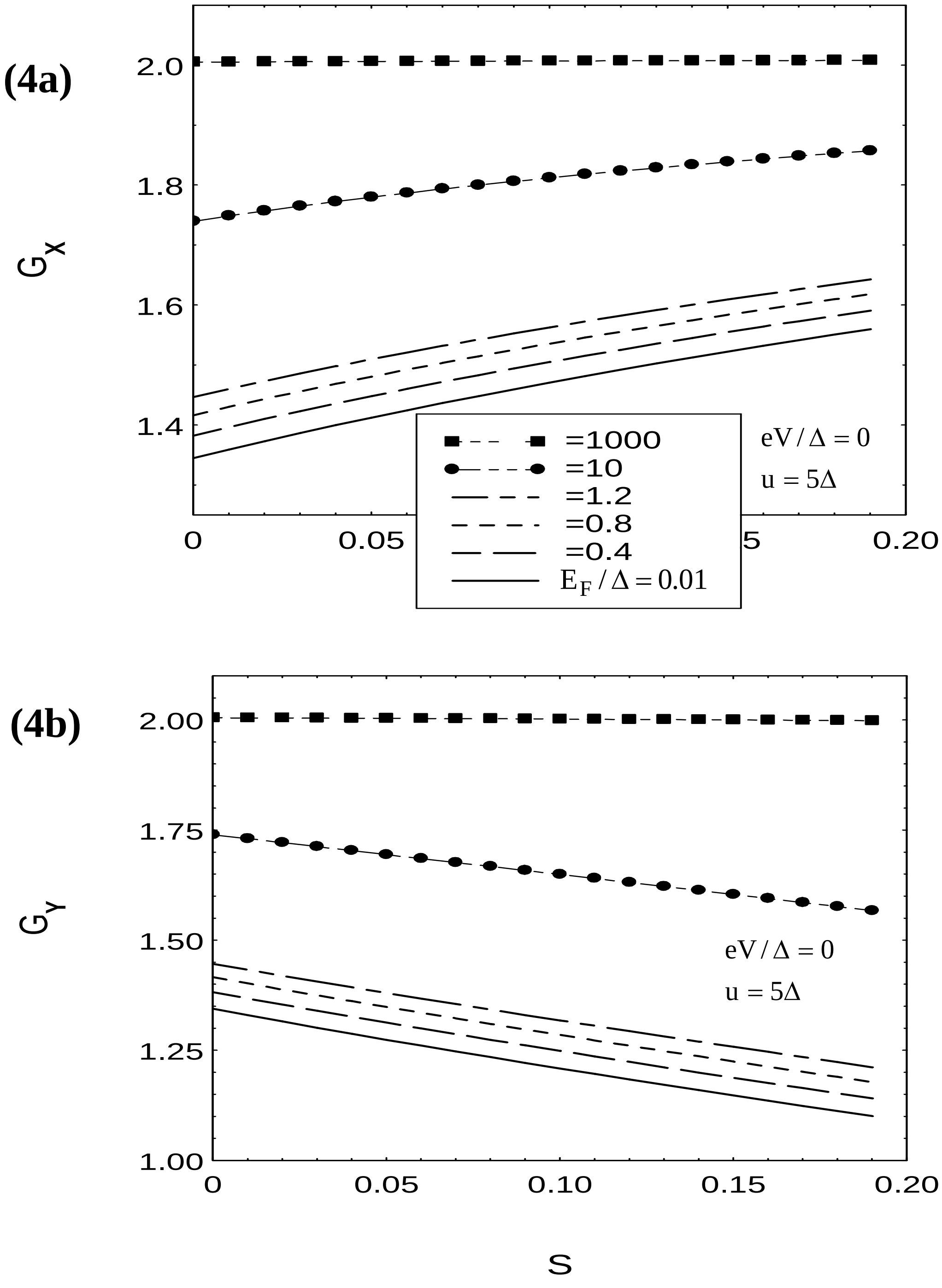

**Figure 4**